\documentclass{blois}

\def\dsttaunu{\bar{B} \rightarrow D^{*} \tau^- \bar{\nu}_\tau}
\def\ddsttaunu{\bar{B} \rightarrow D^{(*)} \tau^- \bar{\nu}_\tau}

\def\dstlnu{\bar{B} \rightarrow D^{*} \ell^- \bar{\nu}_\ell}
\def\ddstlnu{\bar{B} \rightarrow D^{(*)} \ell^- \bar{\nu}_\ell}
\def\dststlnu{\bar{B} \rightarrow D^{**} \ell^- \bar{\nu}_\ell}

\def\Bsig{B_{\rm sig}}
\def\Btag{B_{\rm tag}}
\def\pinu{\tau^- \rightarrow \pi^- \nu_\tau}
\def\rhonu{\tau^- \rightarrow \rho^- \nu_\tau}

\newcommand{\Photo}{}

\begin{document}
\vspace*{4cm}
\title{Recent $\ddsttaunu$ Studies at Belle}

\author{S.~Hirose, for the Belle Collaboration}

\address{Center for Experimental Studies, Nagoya University, Furo, Chikusa, Nagoya, Japan}

\maketitle\abstracts{
  The semitauonic decay $\dsttaunu$ is sensitive to new physics beyond the Standard Model (SM) that has an enhanced coupling to the $\tau$ lepton. In the ratio of branching fractions $R(D^*) = \mathcal{B}(\dsttaunu) / \mathcal{B}(\dstlnu)$, where $\ell^- = e^-$ or $\mu^-$, a 3.3$\sigma$ anomaly was observed. In order to investigate the anomaly further, Belle performed a new $R(D^*)$ measurement using one-prong hadronic $\tau$ decays, which was statistically independent of the previous two measurements. This measurement included the first measurement of the $\tau$ polarization $P_\tau(D^*)$ using the kinematics of the two-body decays. The obtained results, $R(D^*) = 0.270 \pm 0.035 ({\rm stat}) ^{+0.028}_{-0.025} ({\rm syst})$ and $P_\tau(D^*) = -0.38 \pm 0.51 (stat) ^{+0.21}_{-0.16} (syst)$, were consistent both with the SM and the world-average $R(D^*)$. Including this result, the $R(D^*)$ anomaly became 3.4$\sigma$ away from the SM prediction.
}

\section{Introduction}

The decays $\ddsttaunu$ are the semileptonic $B$ meson decays containing a $\tau$ lepton in the final state. These processes are theoretically well studied within the Standard Model (SM), where a virtual $W$ boson mediates the decay at the tree level~\cite{cite:Hwang:2000}. If new physics (NP) beyond the SM exists with a non-universal coupling over the three generation, ratios of branching fractions $R(D^{(*)}) = \mathcal{B}(\ddsttaunu) / \mathcal{B}(\ddstlnu)$, where $\ell^- = e^-$ or $\mu^-$, is modified. Three collaborations, Belle, BaBar and LHCb, have studied the ratios experimentally~\cite{cite:Belle:2015}~\cite{cite:Belle:2016}~\cite{cite:BaBar:2012}~\cite{cite:LHCb:2015}. As of early 2016, the averages of $R(D)$ and $R(D^*)$~\cite{cite:HFLAV:2014} were 1.9$\sigma$ and 3.3$\sigma$ away from the SM predictions~\cite{cite:HPQCD:2015}~\cite{cite:Fajfer:2012}, respectively.

Previously, all the measurements were performed by identifying the $\tau$ lepton from its leptonic decay in order to exploit the presence of one charged lepton in the signal decay for the signal selection. Additionally, hadronic $\tau$ decays can be used to reconstruct signal events. With the full dataset, Belle has performed a new measurement of $\dsttaunu$ using one-prong $\tau$ decays: $\pinu$ and $\rho^- \nu_\tau$~\cite{cite:Belle:2017}. This choice of the $\tau$ decays realizes to investigate the $R(D^*)$ anomaly under the different main background from the previous measurements, where semileptonic decays $\dststlnu$ with excited charmed mesons heavier than $D^*$ are the most important background modes. In addition, using the two-body kinematics of the $\tau$ decay, it is possible to measure the longitudinal polarization of the $\tau$ lepton, $P_\tau(D^*) = (\Gamma^+ - \Gamma^-) / (\Gamma^+ + \Gamma^-)$, where $\Gamma^{+(-)}$ is the decay rate of $\dsttaunu$ with a right-(left-)handed $\tau$ lepton. This variable is sensitive to NP independently of $R(D^*)$~\cite{cite:Tanaka:2013}. The new measurement of $\dsttaunu$ by Belle includes the first experimental study of $P_\tau(D^*)$.

This study is performed based on the dataset accumulated at the center-of-mass $e^+e^-$ collision energy of 10.58~GeV using the KEKB accelerator~\cite{cite:KEKB:2003}. The energy corresponds to the mass of $\Upsilon(4S)$, and $B$ mesons are produced in the process $\Upsilon(4S) \rightarrow B \bar{B}$. The data are recorded by the Belle detector~\cite{cite:Abashian:2002}, which consists of the inner detectors (silicon vertex detector, central drift chamber, time-of-flight counter, aerogel Cherenkov counter and electromagnetic calorimeter), the superconducting solenoid providing a 1.5~T magnetic field, and the $K_L^0$ and muon detector. Out dataset contains 772M $B\bar{B}$ pairs.

\section{Measurement Method}

\subsection{Measurement of $R(D^*)$ and $P_\tau(D^*)$}

The ratio $R(D^*)$ is determined by the yield ratio between the signal mode ($\dsttaunu$) and the normalization mode ($\dstlnu$). It is represented by
\begin{eqnarray}
  R(D^*) &=& \frac{1}{\mathcal{B}_\tau}\frac{\epsilon_{\rm norm}}{\epsilon_{\rm sig}}\frac{N_{\rm sig}}{N_{\rm norm}},
\end{eqnarray}
where $\mathcal{B}_\tau$ denotes the branching fraction of $\tau$, and $\epsilon_{\rm sig (norm)}$ and $N_{\rm sig (norm)}$ are the efficiency and the yield of the signal (normalization) mode, respectively. These quantities are determined separately for different $\tau$ decay modes.

The polarization $P_\tau(D^*)$ is measured by the differential decay rate~\cite{cite:Hagiwara:1990}
\begin{eqnarray}
  \frac{1}{\Gamma} \frac{d\Gamma}{d\cos\theta_{\rm hel}} &=& \frac{1}{2} \left[ 1 + \alpha P_\tau(D^*) \cos\theta_{\rm hel} \right],
\end{eqnarray}
where $\Gamma$ denotes the decay rate of $\dsttaunu$. The coefficient $\alpha$ denotes the $\tau$-mode-dependent sensitivity to $P_\tau(D^*)$; $\alpha = 1$ for $\pinu$ and $\alpha = (m_\tau^2 - 2 m_\rho^2) / (m_\tau^2 + 2 m_\rho^2) \cong 0.45$ for $\rhonu$, where $m_{\tau (\rho)}$ is the $\tau$ lepton ($\rho$ meson) mass. The angle $\theta_{\rm hel}$ is determined as the direction of the $\tau$-daughter $\pi$ or $\rho$ momentum with respect to the direction opposite the momentum of the virtual $W$ boson. However, the complete $\tau$ momentum cannot be measured at Belle due to insufficient kinematic constraints. Here, we exploit the rotation symmetry in the rest frame of $W$. In this frame, we calculate $\cos\theta_{\tau d} = (2 E_\tau E_d - m_\tau^2 - m_d^2) / (2|\vec{p}_\tau||\vec{p}_d|)$, where $E$ and $\vec{p}$ denote the energy and the momentum of $\tau$ and the $\tau$-daughter meson $d = \pi$ or $\rho$, respectively, and $m_d$ is the mass of $d$. The angle $\theta_{\tau d}$ is defined by the momenta of $\tau$ and $d$. Because $\tau$ is produced in the two-body decay, the energy and the magnitude of the $\tau$ momentum is determined only from the momentum transfer squared, $q^2 = (p_{\rm sig} - p_{D^*})$, where $p$ is the four-momenta of the signal $B$ meson ($B_{\rm sig}$) and $D^*$. Although $B_{\rm sig}$ is not fully reconstructed due to two neutrinos, we obtain the $B_{\rm sig}$ momentum from $p_{\rm sig} = p_{e^+ e^-} - p_{\rm tag}$, where $p_{e^+ e^-}$ and $p_{\rm tag}$ denote the four-momenta of the $e^+ e^-$ beam and the counterpart $B$ meson ($B_{\rm tag}$), respectively. In this measurement, we fully reconstruct $B_{\rm tag}$ from its hadronic decay.

Now, the $\tau$ momentum vector is fixed on the cone around the $\tau$-daughter meson momentum with an angle of $\theta_{\tau d}$. Owing to the rotation symmetry of the cone, any direction is kinematically equivalent. An arbitrary direction is therefore selected as a Lorentz boost vector, and we obtain the equation
\begin{eqnarray}
  |\vec{p}_d^{\kern2pt \tau}| \cos\theta_{\rm hel} &=& -\gamma |\vec{\beta}| E_d + \gamma |\vec{p}_d| \cos\theta_{\tau d},
\end{eqnarray}
where $|\vec{p}_d^{\kern2pt \tau}| = (m_\tau^2 - m_d^2) / (2 m_\tau)$ is the $\tau$-daughter momentum in the rest frame of $\tau$, and $\gamma = E_\tau / m_\tau$ and $|\vec{\beta}| = |\vec{p}_\tau| / E_\tau$. Solving this equation, the value of $\cos \theta_{\rm hel}$ is obtained. 

\subsection{Event Reconstruction}

The event selection starts with reconstructing $\Btag$ from one of the 1104 hadronic decay chains. The NeuroBayes-based multivariate analysis technique~\cite{cite:Feindt:2011} is employed. Good quality $\Btag$ candidates are selected based on the beam-constraint mass $M_{\rm bc} = \sqrt{E_{\rm beam}^{*2} - |\vec{p}_B^{\kern2pt *}|^2}$, where $E_{\rm beam}^*$ and $\vec{p}_B^{\kern2pt *}$ are the beam energy (5.29~GeV) and the three momentum of the $\Btag$ candidate, respectively, as well as the single NeuroBayes output classifier $\mathcal{O}_{\rm NB}$. Our requirements retain 90\% of correctly reconstructed $B_{\rm tag}$ candidates while rejecting 70\% of misreconstructed candidates.

Using the remaining particles, we next reconstruct the $D^*$ meson as a daughter of $\Bsig$. Reconstructed $D^*$ candidates are then combined with pions or $\rho$ meson candidates properly in terms of the charge. The $\rho$ meson candidates are reconstructed from the decay $\rho^- \rightarrow \pi^- \pi^0$. For the normalization mode, we require the existence of one $e^-$ or $\mu^-$ instead of $\pi^-$ or $\rho^-$. Finally, we select events with no extra charged track and $\pi^0$ candidate not used for the reconstruction of $\Btag$ and $\Bsig$.

Details of the event reconstruction are discussed in Ref.~\cite{cite:Belle:2017}.

\section{Result}

\begin{figure}[t!]
  \begin{minipage}{0.5\textwidth}
    \centering
    \includegraphics[angle=90,width=6cm]{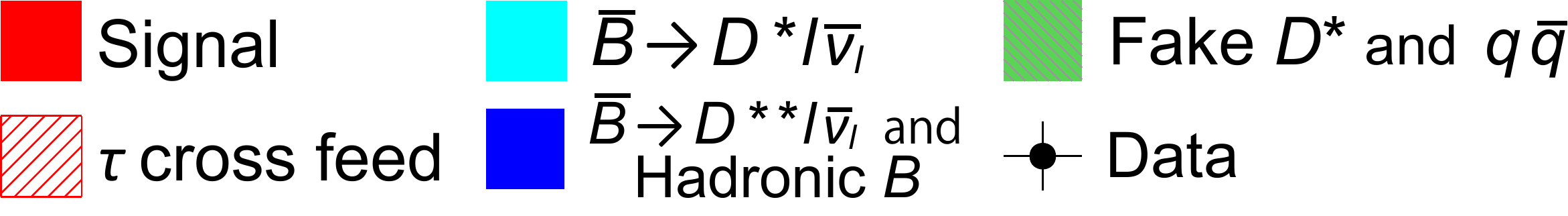}
    \includegraphics[width=7.5cm]{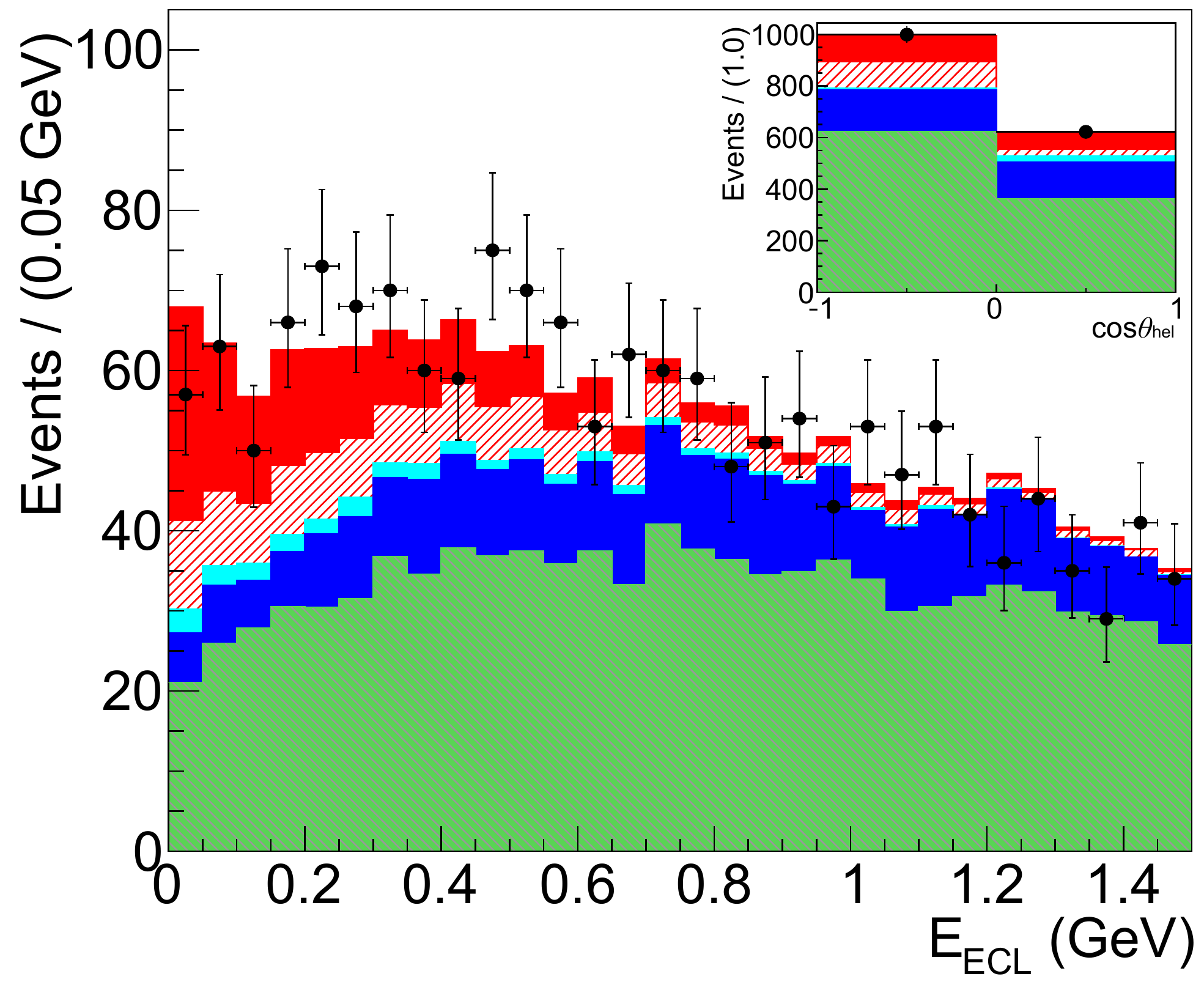}
    \caption{Fit results to the signal sample. The main panel and the sub panel show the $E_{\rm ECL}$ and the $\cos\theta_{\rm hel}$ distributions, respectively. The red-hatched ``$\tau$ cross feed'' contains $\dsttaunu$ signal events originating from $\tau$ decays different from the reconstruction channels.}
    \label{fig:fit-result}
  \end{minipage}
  \begin{minipage}{0.5\textwidth}
    \centering
    \includegraphics[width=6cm]{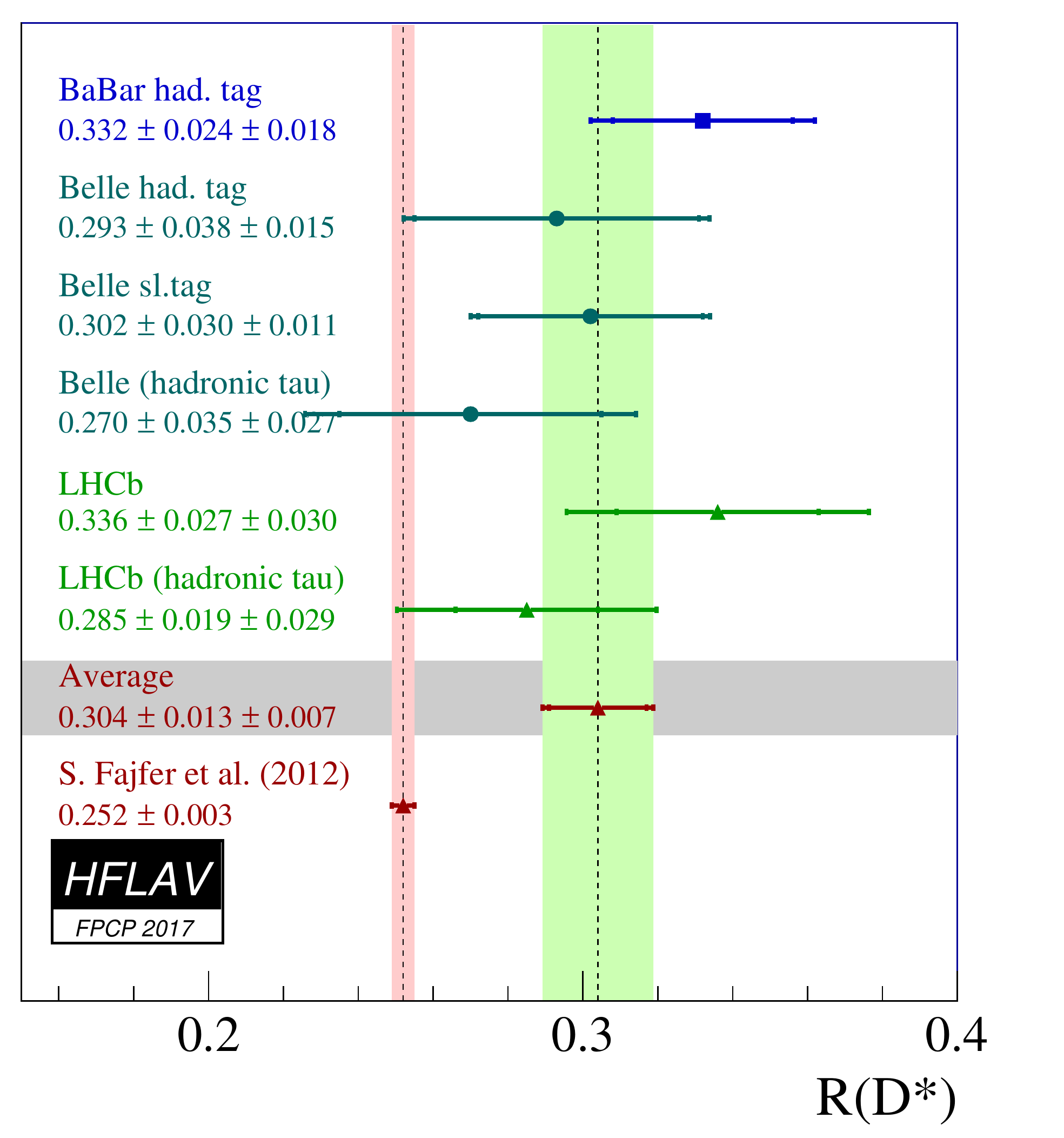}
    \caption{Summary of the $R(D^*)$ measurements as of FPCP 2017.}
    \label{fig:RDst-summary}
  \end{minipage}
\end{figure}

Signal extraction is performed by the two-step fit. First, a fit to the normalization events is performed based on the missing-mass squared $M_{\rm miss}^2 = (p_{e^+ e^-} - p_{\rm tag} - p_{D^*} - p_\ell)^2$, where $p_\ell$ denotes the four-momentum of $\ell$. Since there is only one neutrino in the normalization mode, the distribution of $M_{\rm miss}^2$ peaks around 0~GeV$^2$. After determining the normalization yield, we perform a fit to the signal events using $E_{\rm ECL}$, which is the sum of the cluster energies on the electromagnetic calorimeter that are not used for the event reconstruction. While the signal events tend to have $E_{\rm ECL}$ close to 0~GeV, the background events have larger values due to additional physical photons. The $E_{\rm ECL}$ has advantages for the signal yield extraction in terms of its small correlation to $P_\tau(D^*)$ and good signal separation from background processes. In the fit, the $P_\tau(D^*)$ is determined from two bins of $\cos\theta_{\rm hel}$.

Figure.~\ref{fig:fit-result}~\cite{cite:Belle:2017} shows the fit results to the signal sample. From the fit, we obtain
\begin{eqnarray}
  R(D^*)   &=& 0.270 \pm 0.035 ({\rm stat}) ^{+0.028} _{-0.025} ({\rm syst}),\\
  P_\tau(D^*) &=& -0.38 \pm 0.51 ({\rm stat}) ^{+0.21} _{-0.16} ({\rm syst}).
\end{eqnarray}
The systematic uncertainty mainly arises from hadronic and semileptonic decays of $B$ mesons, and the statistics of the Monte Carlo simulated sample; details are described in Ref.~\cite{cite:Belle:2017}. The signal significance is 7.1$\sigma$, and therefore our measurement has achieved the first observation of $\dsttaunu$ only with $\tau$ decays. The region $P_\tau(D^*) > +0.5$ is excluded at the 90\% confidence level, which is the first result of $P_\tau(D^*)$ in $\dsttaunu$.

\section{Current Situation of $R(D^*)$}

In addition to our new result, LHCb has performed a measurement of $R(D^*)$ using three-prong hadronic $\tau$ decays~\cite{cite:LHCb:2017}. As shown in Fig.~\ref{fig:RDst-summary}~\cite{cite:HFLAV:2016}, the current world-average $R(D^*) = 0.304 \pm 0.013 ({\rm stat}) \pm 0.007 ({\rm syst})$ is 3.4$\sigma$ away from the SM prediction~\cite{cite:HFLAV:2016}. Including $R(D)$, the overall discrepancy reaches 4.1$\sigma$. In order to settle this puzzle, further $\ddsttaunu$ studies using high statistics data at Belle~II are essential.

\section{Conclusion}

The decays $\ddsttaunu$ are interesting $B$ decays in terms of their sensitivities to NP coupling to the $\tau$ lepton. Belle has performed a new $\dsttaunu$ measurements, which results in the first observation of the $\dsttaunu$ signal using only hadronic $\tau$ decays, and the first measurement of $P_\tau(D^*)$. The obtained result is consistent with the SM predictions. The world averages of $R(D^{(*)})$ show the 4.1$\sigma$ discrepancy from the SM predictions. This is an important topic to be further investigated with high precision of Belle~II. 
\section*{Acknowledgments}

This work was partially supported by JSPS Grant-in-Aid for Scientific Research (S) ``Proving New Physics with Tau-Lepton'' (No.~26220706).

\end{document}